# Electrically-driven optical antennas

Johannes Kern[1,2,*], René Kullock[1,*], Jord Prangsma[3], Monika Emmerling[4], Martin Kamp[4] and Bert Hecht[1,#]

[1]Nano-Optics and Bio-Photonics Group, Department of Experimental Physics 5, University of Würzburg, Am Hubland 97074 Würzburg, Germany.
[2]Ultrafast Solid-State Quantum Optics and Nanophotonics Group, Physikalisches Institut, Westfälische Wilhelms-Universität Münster, Wilhelm-Klemm-Str. 10 48149 Münster, Germany.
[3]Nanobiophysics and Optical Sciences, MESA+ Institute for Nanotechnology, University of Twente, P.O. Box 217, 7500 AE, Enschede, The Netherlands.
[4]Technische Physik and Wilhelm-Conrad-Röntgen-Research Center for Complex Material Systems, University of Würzburg, Am Hubland 97074 Würzburg, Germany.
*These authors contributed equally to this work.

[#]e-mail: hecht@physik.uni-wuerzburg.de

**Unlike radiowave antennas, optical nanoantennas so far cannot be fed by electrical generators. Instead, they are driven by light[1] or via optically active materials in their proximity[2]. Here, we demonstrate direct electrical driving of an optical nanoantenna featuring an atomic-scale feed gap[3]. Upon applying a voltage, quantum tunneling of electrons across the feed gap creates broadband quantum shot noise. Its optical frequency components are efficiently converted into photons by the antenna. We demonstrate that the properties of the emitted photons are fully controlled by the antenna architecture, and that the antenna improves the quantum efficiency by up to two orders of magnitude with respect to a non-resonant reference system. Our work represents a new paradigm for interfacing electrons and photons at the nanometer scale, e.g. for on-chip wireless data communication, electrically driven single- and multiphoton sources, as well as for background-free linear and nonlinear spectroscopy and sensing with nanometer resolution.**

Radio- and microwaves can be generated by currents that oscillate within antennas driven by high-frequency voltage sources reaching up into the 100 GHz regime. Sources for optical radiation are traditionally based on transitions between quantum states since conventional electrical circuits are unable to generate oscillating currents at optical frequencies at around 300 THz[4]. As a result, the well-developed and powerful concepts of antenna theory are difficult to apply to optical radiation opposite to what Feynman had anticipated[5]. However, already in 1976 it was shown that visible light can be generated through electron tunneling in metal-insulator-metal junctions[6]. Soon after, it was proposed that such light emission is caused by quantum shot noise resulting in voltage-dependent broad-band current fluctuations reaching into the optical frequency range[7]. Further experimental observations of light emission from different tunnel contacts supported this idea[8–10], however, only recently the quantum shot noise picture has been proven conclusively[11]. Here we exploit quantum shot noise to generate optical-frequency current oscillations within an antenna gap creating for the first time an electrically-driven optical antenna.

In order to realize an electrically driven optical antenna the challenge is to implement a lateral tunnel junction in the feed gap of an electrically connected optical antenna on an insulating transparent substrate[12]. To this end we combine top-down focused-ion beam structuring of single-crystalline gold flakes on glass substrates[13] and AFM nano-manipulation of monodisperse spherical gold nanoparticles covered with a ligand shell[3]. When pushed into a tailor-made antenna gap, the molecular spacer layer covering the nanoparticles leads to a comparatively stable atomic-scale gap (cf. Fig. 1a).

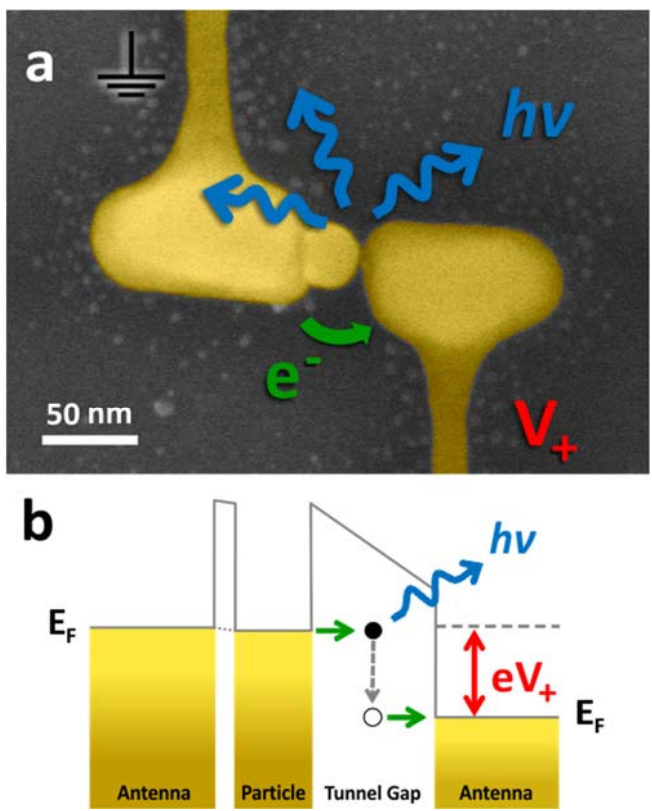

**Figure 1 | Electrically-driven optical antenna. a,** Electron micrograph of a lateral tunnel structure: an electrically-connected single-crystalline gold nanoantenna loaded with a coated gold nanoparticle on a glass substrate. $V_+$: applied voltage, $e^-$: electron flow, and $h\nu$: light emission. **b,** Tunnel gap with applied voltage.

For applying a DC voltage between the antenna arms we connected them at minimum optical field points with single-crystalline gold nanowires, which leaves the antenna resonance unchanged[12]. These connectors extend for some micrometers before being attached to macroscopic contact pads (Supplementary Information). We have fabricated a large number of devices and respective SEM images suggest that typically the particle is not exactly centered in the antenna gap, in particular after a voltage has been applied. Therefore, since tunnel resistances increase exponentially with gap size, for most devices the applied voltage drops over a single junction as sketched in Fig. 1b. Note that, due to surface self-diffusion[14] and the high DC fields, gold atoms will rearrange around the gap which leads to an increasing current over time. Nevertheless, in our considered time frames the current was stable (Supplementary Information).

In Fig. 2a an exemplary current vs. voltage (I-V) curve is plotted. With increasing voltage the current first increases linearly and later superlinearly. The corresponding Fowler-Nordheim representation (inset) is fully compatible with a dominating single tunnel junction showing direct tunneling for low and field emission at high voltage as well as a single dip at the transition voltage $V_T$. A fit to the data using a single-barrier model[15] is consistent with a 1.3 nm gap and a barrier height of 2.6 eV (Supplementary Information). Such reduced barrier heights are commonly observed at ambient conditions[16]. Many reproducible I-V curves can be recorded for such a junction although they exhibit current fluctuations typical for tunnel experiments at the nanoscale (Supplementary Information).

For sufficiently high voltages the tunneling of electrons is accompanied by emission of visible photons. The emitted photons are collected through a high-numerical aperture microscope objective and directed towards a sensitive detection system (see methods). The number of photons increases linearly with the tunneling current. A typical far-field emission spot is overlaid with an SEM micrograph of the antenna area and depicted in Fig. 2b. The recorded Airy pattern is co-localized with the antenna and exhibits a FWHM of ~350 nm, close to the diffraction limit. Hence, the device acts as an electrically-driven subwavelength photon source.

The spectrum of the emitted light is expected to depend on the applied voltage and the antenna resonance. Therefore we have characterized the antenna resonance by white-light dark-field scattering spectroscopy (see methods). The scattering spectrum shows that the particle-loaded antenna has two resonances: one at lower energies around 760 nm and a second one at higher energies around 590 nm. Due to its stronger field confinement and enhancement, only the lower energy resonance significantly enhances light emission (Supplementary Information). The influence of the applied voltage on the light emission is presented in Fig. 2c, where we plot electroluminescence (EL) spectra for applied voltages ranging from 1.7 to 2.0 V. Each spectrum features a peak and a high-energy cut-off that both blue-shift with increasing voltage while the peak also gains intensity. A similar blue-shift of the photon spectrum for increasing applied voltage was observed before in MIM structures[17]. Note that no further shift is observed for voltages sufficiently larger than the antenna resonance energy and that the electroluminescence spectrum closely resembles the antenna scattering spectrum under these operating conditions (Supplementary Information). In general, the spectrum of emitted photons can be understood by considering the spectral density $C(\omega)$ of the temporal fluctuations of the tunneling current[18], which can be expressed as $C(\omega) \sim (eV - \hbar\omega)$ with spectral components at frequency $\omega$ that reach well beyond the optical frequency range depending on the applied voltage $V$. To model the interaction of the fluctuating tunnel current with the antenna resonance we assume that it can be represented by a sum of single-frequency dipoles located in the tunnel junction with amplitudes according to $C(\omega)$. Photon creation is considered a two-step process. In a first step, the dipoles excite the antenna resonance with finite efficiency. In a second step, the antenna radiates into the far-field and the emission spectrum is determined by the far-field response. Our single-crystalline antennas exhibit relatively narrow resonances and it has been shown that under such circumstances near- and far-field response are almost identical[19]. Consequently, in good approximation, it is possible to model the electroluminescence spectrum by multiplying the spectral density of the current fluctuations with the experimentally determined far-field scattering spectrum of the antenna, which includes both the excitation spectrum and the emission characteristics. The solid lines in Fig. 2c have been calculated according to this model for different applied voltages $V$ and are in excellent agreement with the respective electroluminescence spectra. Most importantly, the model predicts the observed blue-shift of the electroluminescence and the fact that the energy of emitted photons cannot be larger than the applied voltage considering only linear processes (quantum cut-off)[20]. The fact that the onset of photon emission at high energies appears at the applied voltage confirms the effective single-junction tunneling model.

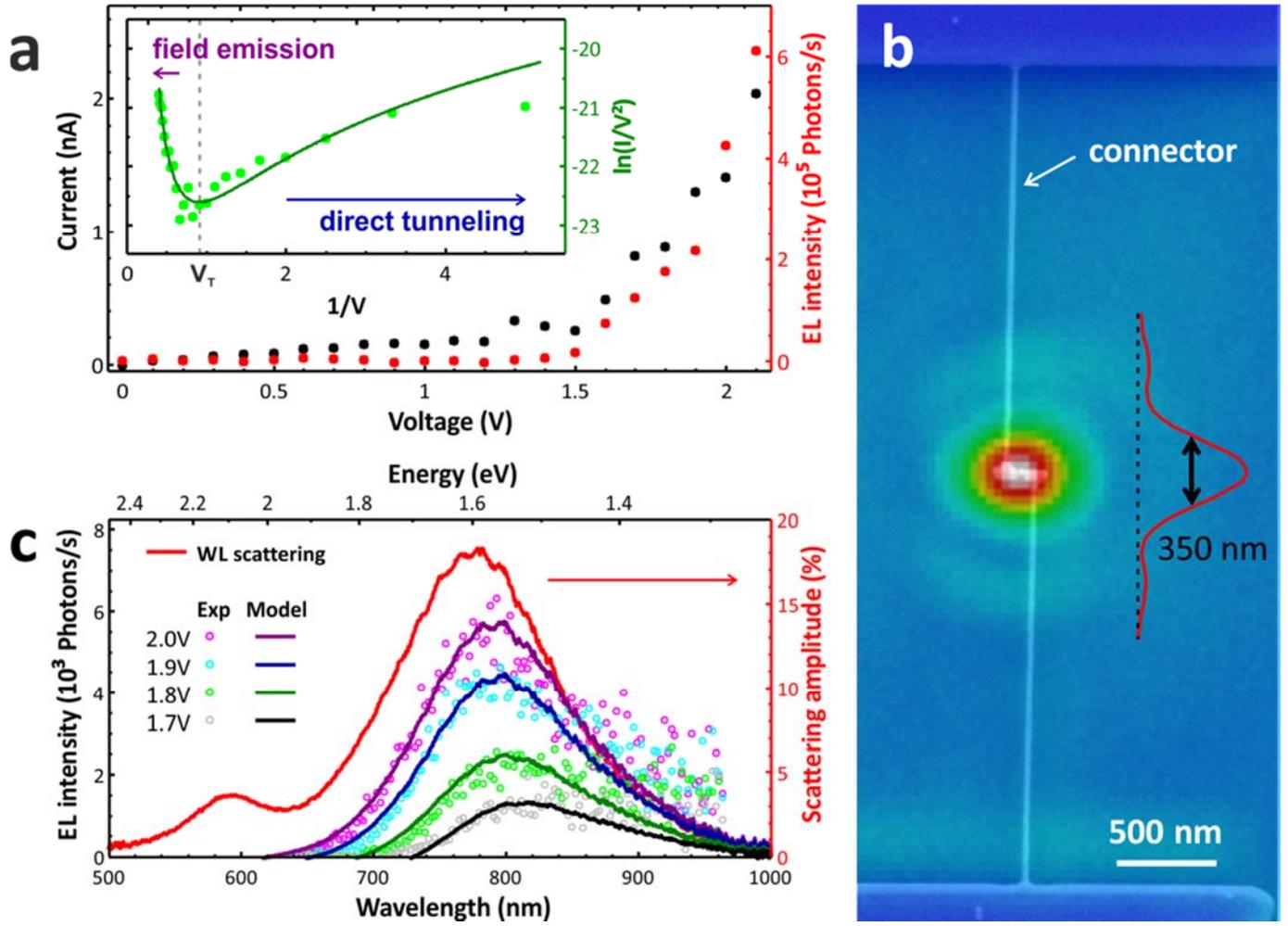

**Figure 2 | Electro-optical characterization. a,** Current vs. voltage (I-V) plot of a connected antenna with tunnel gap. For applied voltages above 1.5 V light is detected whose intensity grows linearly with the current. Inset: Fowler-Nordheim representation of the same I-V data and fit (solid line) using the Simmons model (with image charges, $s$=1.27 nm, $\varphi_{eff}$=2.56 eV, $w$=2.14 nm, $\varepsilon_{CTAB}$=1.435, $I_{offset}$=50 pA, $I_{leakage}$=0, see Supplementary Information). **b,** Electron micrograph of a structure superimposed with the subwavelength emission spot showing an Airy pattern with a FWHM of 350 nm. Weak background illumination was used to visualize the outline of the structure. **c,** Electroluminescence spectra for various applied voltage (open symbols). Solid red line: scattering spectrum of an unbiased antenna. Solid colored lines represent calculated electroluminescence spectra obtained according to the model described in the text. A global scaling factor was used to match the experimental data.

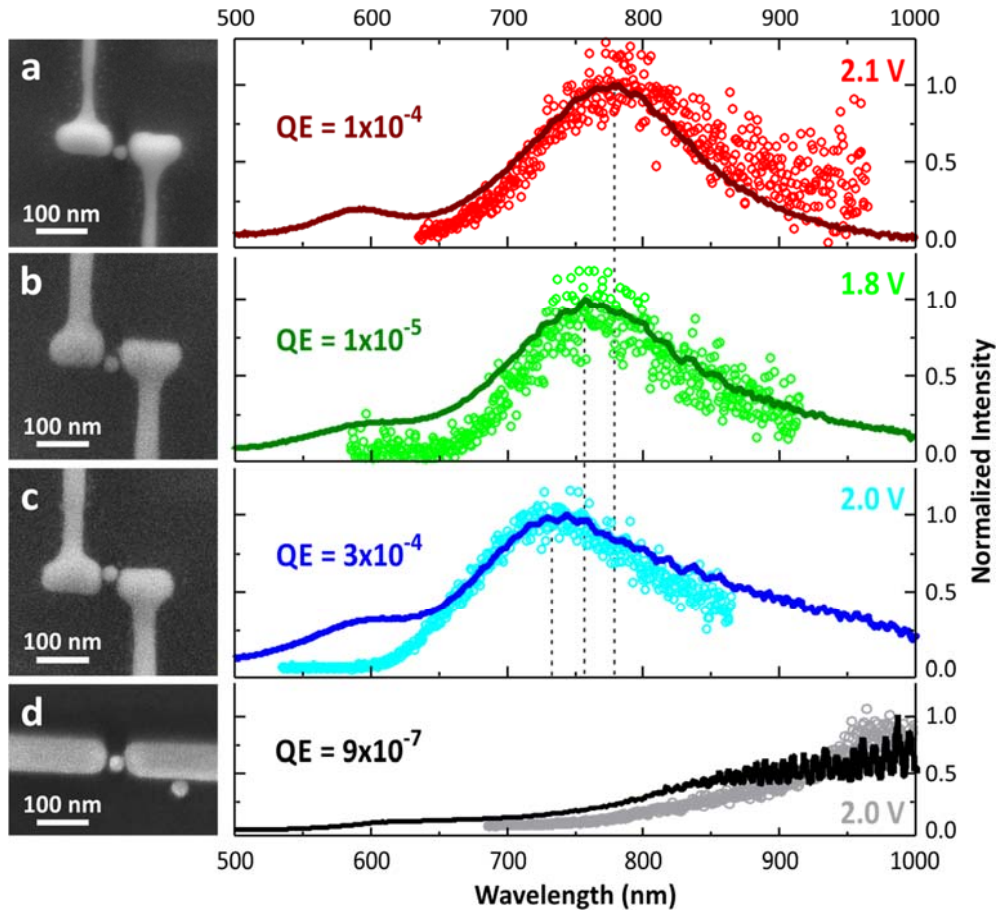

**Figure 3 | Tunability and efficiency.** Electroluminescence (open circles) and scattering spectra (solid lines) for a-c different antenna geometries and d a non-resonant several micrometer long wire. Left: Electron micrographs of the corresponding structures. The external quantum efficiencies (QE) and applied voltages are stated.

We next demonstrate that the observed electroluminescence spectrum can be modified by changing the antenna geometry. For comparison we also investigate a non-resonant reference structure consisting of two several-μm long wires between which a tunneling gap was prepared. Fig. 3a-d depict electron micrographs of the corresponding structures (left) and normalized scattering as well as matching electroluminescence spectra recorded at sufficiently high voltage (right). For the resonant antennas pronounced peaks are observed at wavelengths of 732, 763 and 783 nm, respectively. The non-resonant wire does not show a peak but the signal slowly increases towards longer wavelengths as it is dominated by the behavior of the current spectral density.

For a quantitative comparison between resonant and non-resonant systems, we have determined the external quantum efficiency of the electron to photon conversion (Supplementary Information) by dividing the number of emitted photons by the number of electrons passing the tunnel junction (see also Fig. 3). The quantum efficiencies for the resonant antennas range from $10^{-5}$ to $10^{-4}$. On the other hand the obtained efficiency for the non-resonant wire is lower than $10^{-6}$. Consequently, the best antenna exhibits a two orders of magnitude larger efficiency compared to the non-resonant wire system and a one order of magnitude larger efficiency than typical values reported for STM light emission at ambient conditions[21].

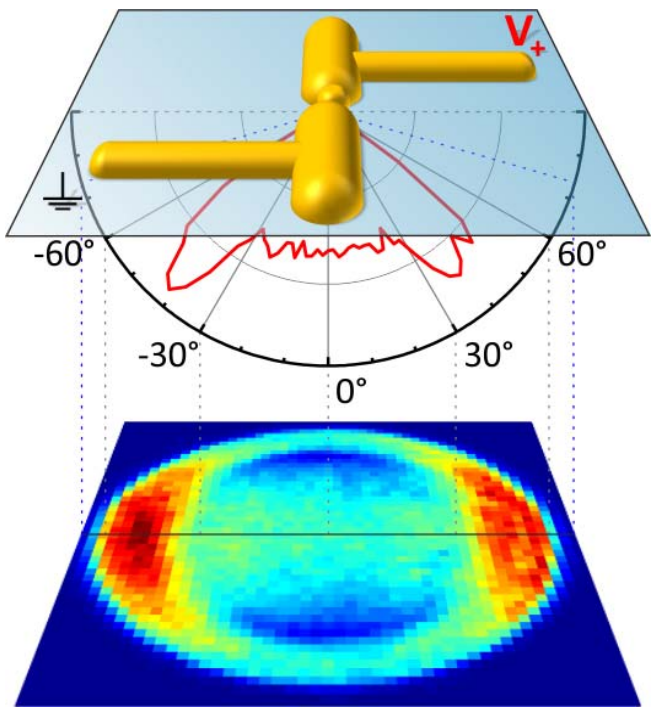

**Figure 4 | Radiation pattern.** Experimentally obtained radiation pattern of an electrically-driven antenna (1.8 V, ~2.5 nA for 1 s) revealing a dipole-like pattern.

The antenna also governs the angular radiation pattern of the emitted photons. To obtain the radiation pattern we image the Fourier plane of the microscope objective (Fig. 4). The radiation pattern shows two lobes close to the critical angle similar to the pattern of a dipole in proximity to a glass surface (Supplementary Information). This is expected for the radiating mode as the charge oscillations on the antenna arms resemble that of two in-phase sub-diffraction dipoles.

Furthermore, the polarization of the emitted light measured after the microscope objective exhibits a 10:1 ratio along the antenna axis (Supplementary Information). The obtained value is limited by the depolarization introduced by the high-NA collection optics.

In conclusion, we have devised a new type of electrically-driven subwavelength photon source based on an electrically-connected plasmonic nanoantenna. The antenna wires not only support a resonant mode at optical frequencies, but are also used to apply a voltage across the tunnel junction that separates the antenna wires. In contrast to semiconductor or organic light-emitting devices, our device consists of a single material and its properties are not limited by an internal electronic structure but rather depend on the antenna architecture. The planar geometry results in a voltage-independent gap, well-defined resonances, easy external access to the tunnel gap as well as possibilities for extending the antenna[2,22] and electrode design[23,24]. It therefore provides large potential for fundamental studies as well as applications. Since the fluctuations in the tunneling current constitute a broadband source, the electroluminescence can be tuned to the blue as well as to the infrared spectral region e.g. by using shorter and longer antennas, respectively.

Possible applications lie in the field of electronic and photonic hybrid circuitry. Large bandwidth optical transistor operation may be realized by means of photon-assisted tunneling[25] or by integrating molecular switches[26] into the tunnel junction. For quantum communication applications, it would be exciting to imprint the fermionic nature of electrons onto photons in order to realize a source of single photons[27,28]. Towards practical implementations it will be important to further improve the stability as well as the efficiency of the devices. The tunneling junction could be further stabilized by the evaporation of an insulator or by exploiting a covalent bond of the molecular layer around the gold particle to the antenna arms, and the quantum efficiency could be improved by exploiting cross-conjugated molecules suppressing elastic tunneling by quantum interference[29] or by utilizing antenna geometries featuring higher field enhancements.

**Methods**

*Fabrication*

Single-crystalline gold flakes are grown on a borosilicate coverslip in a wet-chemical process via reduction of chloroauric acid $HAuCl_4$ in ethylene-glycol[30]. Flakes are then transferred to an electrode structure. The electrode structure has been defined by optical lithography and consists of a 20 nm chromium adhesion layer and a 100 nm gold layer. Nanoantennas are fabricated from the single-crystalline gold flakes by focused-ion-beam milling (Helios Nanolab, FEI Company, Oregon, USA) using an acceleration voltage of 30 kV and a beam current of 1.5 pA. A diluted solution of gold nanoparticles (A11C-30-CTAB-1, Nanopartz, Loveland, USA) is drop-casted onto the antennas. In order to remove excess of the molecular surfactant, the sample has been rinsed in 60° C warm ultrapure water and exposed to oxygen plasma. Particles are manipulated by an atomic force microscope (Veeco Dimensions 3100) using a silicon cantilever (OMCL-AC240TSE, Olympus).

*Optical Characterization*

A halogen lamp coupled into a multi-mode fiber serves as an excitation source for the dark-field white light scattering measurement. The light is focused into the back focal plane of an oil-immersion microscope objective (Plan-Apochromat, 63x, NA =1.40, Zeiss) in order to illuminate the sample with a parallel light beam. In the detection path direct reflection is blocked by a circular beam block, such that only the scattered light is collected. The scattered light is dispersed by a spectrometer (Shamrock 303i, 80 lines/mm blazing at 870 nm) and detected by an electron-multiplied charge-coupled device (iXon A-DU897-DC-BVF, Andor).

*Electrical Characterization*

The electrodes are contacted by copper-beryllium probe needles (Semprex Corp., Cambell USA) which are mounted onto a probehead (DPP220, Cascade Microtech, Beaverton, USA). Current-voltage characteristics are recorded by a source meter unit (Keithley 2601A, Keithley Instruments Inc., Cleveland, USA) using a guard and noise shield.

*Electro-optical measurements*

Voltages are applied by the source meter unit (Keithley 2601A, Keithley Instruments Inc., Cleveland, USA) and electroluminescence is collected by an oil-immersion microscope objective (Plan-Apochromat, 63x, NA =1.40, Zeiss) and detected through a spectrometer (Shamrock 303i, 80 lines/mm blazing at 870 nm or mirror) by an electron-multiplied charge-coupled device (iXon A-DU897-DC-BVF, Andor). Typical integration times of the camera are 100 ms. Source meter unit and EMCCD camera have been synchronized by a LabVIEW program in order to allow for a correlated data analysis.

**Supplementary Information** is found at the end of this document

**Acknowledgements**
We gratefully acknowledge experimental support by S. Großmann and financial support of the VW-foundation (Grant I/84036) and the DFG (HE 5618/4-1).

**Author Contributions**
J.K, R.K, J.P. & B.H. conceived the experiment. J.K, R.K & J.P. designed the antennas. J.K & M.E. designed and fabricated the electrode structure. R.K. grew the gold flakes and transferred them. J.K. & R.K. milled the structures and performed the particle pushing. M.K. supervised the FIB fabrication. R.K. programmed experiment control and data acquisition software. J.K. & R.K. constructed the experiment, performed the measurements and analyzed the data. J.K. performed the FDTD simulations. J.K, R.K., J.P. & B.H. co-wrote the manuscript with input from all authors.

# Supplementary Information

## Electrically-driven optical antennas

Johannes Kern, René Kullock, Jord Prangsma, Monika Emmerling, Martin Kamp and Bert Hecht

## Table of Contents

# S1.  Fabrication

The fabrication of electrically driven optical antennas consists of multiple steps to ensure good electrical contact, precise nanoantenna structuring and an atomic-scale gap formation.

## Electrode Structure

Direct contacting of single-crystalline gold flakes using e.g. needles positioned by micromanipulators does not result in stable electrical contacts. In order to achieve a stable contact we therefore employ a softer Au/Cr electrode structure as indicated in Figure S1.1. The electrode structure consists of multiple electrode fingers allowing us to fabricate and investigate several electrically connected antennas on a single gold flake.

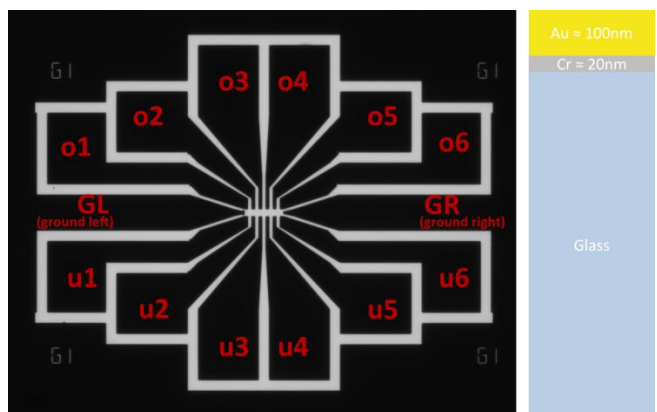

**Figure S1.1 | Layout of the electrode structures. Left:** Top view of the structure with two ground electrodes and twelve (o1-o6, u1-u6) contact electrodes colored in black. **Right:** Schematics of the stack composition.

The electrode structure was fabricated by optical lithography using a negative mask, fabricated by e-beam lithography. A thin adhesion layer (AR 300-80, ALLRESIST GmbH, Strausberg, DE) is necessary, which was spin-coated onto cleaned microscope cover slips (#1, Gerhard Menzel GmbH, Saarbrücken, DE) and afterward baked at 95° C for 60 s. Next a 1.3 µm thick image reversal resist (AR-U4030, ALLRESIST GmbH, Strausberg, DE) was spin-coated and afterwards hardened at 95° C. The resist was exposed through the negative mask for 13 s, reversal baked on a hotplate (180 s, 120° C) and afterwards flood exposed (maskless, 60 s, dose doubled). For development AR-300-35 (ALLRESIST GmbH, Strausberg, DE) solved in water (1:1) was applied for 20 s. The samples were subsequently rinsed with water and plasma cleaned (200 W, 60 s, 50% oxygen). Afterwards a 20 nm chromium adhesion layer and a 100 nm layer of gold were evaporated with electron beam physical vapor deposition at a rate of 0.5 Å/s. The remains were lifted off by putting the samples in methylpyrrolidone, heating it up to 80° C and applying an ultrasonic cleaning (130 kHz, 120 s). Finally, the samples were rinsed with acetone.

## Flakes Growth & Transfer

Single-crystalline gold flakes largely facilitate the fabrication of precise plasmonic nanostructures using focused-ion beam milling[S1]. Gold flakes were grown directly on glass substrates[S2] based on a recipe by Guo and coworkers[S3]. Lateral large (>100 µm) but thin (50-80 nm) flakes were then selected and transferred to the electrode structure (see Figure S1.2) using an ethanol immersion and micromanipulators. Typical resulting contact resistances between flake and electrode were below 1 Ω.

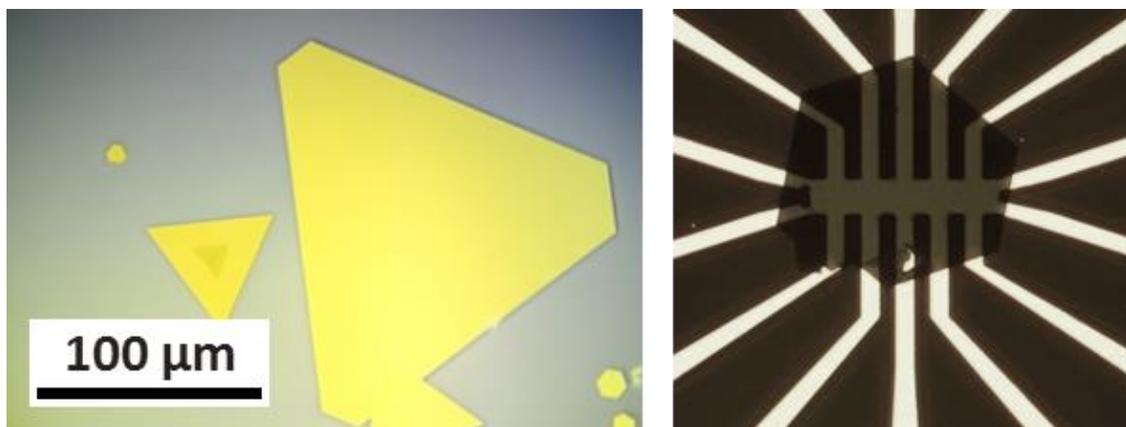

**Figure S1.2 | Flake growth and transfer. Left:** Optical reflectance image recorded using a color CCD (see section S2, optical imaging and spectra) of single-crystalline flakes on glass. **Right:** transmission image of an individual flake placed on an electrode structure.

## FIB Structuring

The transferred flakes were structured using focused-ion beam milling (FIB, Helios Nanolab 600). Different apertures were used to define the ion current for coarse and medium resolution milling. For high-resolution structuring a 16 µm aperture (9.7 pA) and an acceleration voltage of 30 kV was used. Typical dimensions of the resulting antennas are 50-80 nm height, 60 nm width, 240-300 nm length, 20-30 nm gap and a connector width/height of 30-40 nm. Directly after the structuring the antennas were characterized via optical microscopy (images, spectra) and scanning electron microscopy (SEM, see Figure S1.3).

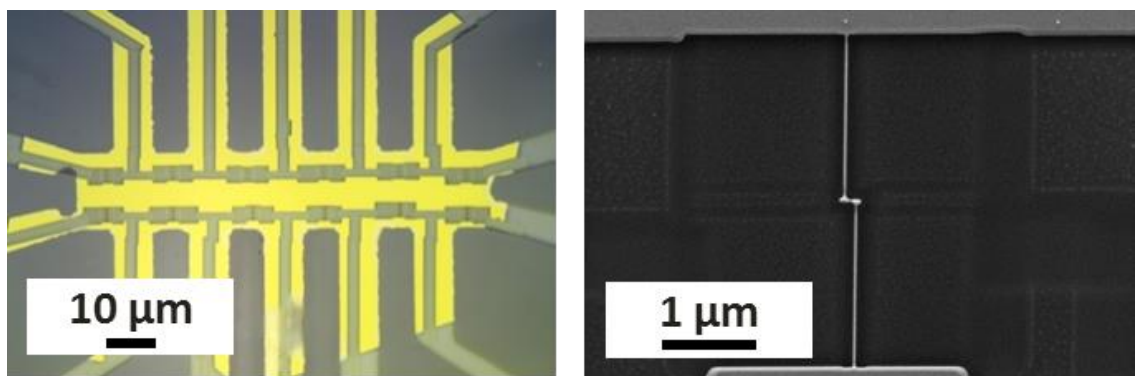

**Figure S1.3 | FIB structuring. Left:** Optical reflectance image of a structured flake (bottom view) and **Right:** SEM zoom-in to a single antenna structure**.**

## Particle Dropping and Pushing

In order to fabricate the tunneling gaps, CTAB decorated gold particles (diameter 30 nm, A11C-30-CTAB-1, Nanopartz, Loveland, USA) dissolved in water were drop-casted onto the structured samples. The dilution was adjusted to result in a surface density of roughly ~1 particle/µm². After the drop evaporated, the whole sample was washed for 30 s in warm water (60° C), dried and exposed to oxygen-plasma for 30 s to remove remaining CTAB. Then, SEM images were acquired to identify suitably arranged particles and structures.

Finally, nanomanipulation of the gold particles was performed using an AFM (Veeco Dimension Icon) in both feedback and non-feedback mode. On the one hand, particles were pushed into the antenna in order to form an atomic-scale gap (Figure S1.4). On the other hand, further surrounding particles were pushed far away from the antenna to prevent parasitic scattering in white-light scattering experiments . Directly after nanomanipulation the antennas were characterized by optical microscopy (images, spectra) and the arrangement was checked by SEM (only recording rough overview images to prevent damage or carbon deposition).

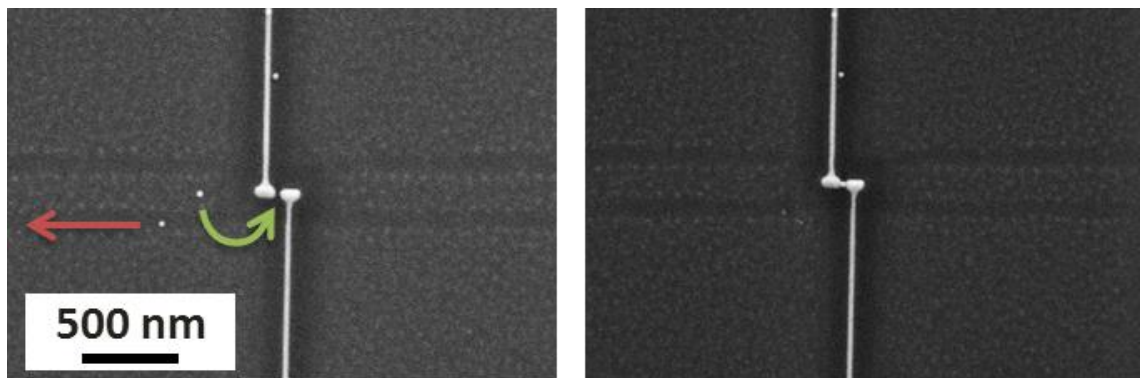

**Figure S1.4 | AFM pushing. Left:** SEM image of drop-casted nanoparticles close to an antenna and **Right:** SEM of the same structure after nanomanipulation.

# S2.  Characterization

## SEM imaging

SEM images were acquired using a Zeiss Ultra Plus electron microscope equipped with a gas injection needle. Oxygen was supplied locally while scanning, to reduce carbon deposition on the structures. Fast overview images (magnification <200k) were acquired at several stages of the fabrication process but slow high-resolution measurements (magnification >200k, e.g. Figure 1 in text) were only conducted after electroluminescence measurements to exclude carbon contamination of the tunneling junction.

## Optical images and spectra

Optical images (Figure S1.2 and S1.3) were recorded using an Olympus IX70 microscope equipped with an 100x NA 1.45 objective, a Halogen (transmission) or Mercury-arc-lamp (reflection mode), and a CCD camera (Canon EOS 100D).

White-light dark field scattering spectra were recorded using a homebuilt setup (Figure S2.1). A halogen lamp coupled into a multi-mode fiber served as an excitation source. The light is focused into the back focal plane of an oil-immersion microscope objective (Plan-Apochromat, 63x, NA =1.4, Zeiss) in order to illuminate the sample with a parallel light beam. In the detection path direct reflection is blocked by a circular beam block, such that only the scattered light is collected. The scattered light is dispersed by a spectrometer (Shamrock 303i, 80 lines/mm blazing at 870 nm) and detected via an electron-multiplied charge-coupled device (EMCCD, iXon A-DU897-DC-BVF, Andor). A stitching method ("step 'n glue" from Andor) was used to cover a sufficiently wide spectral range.

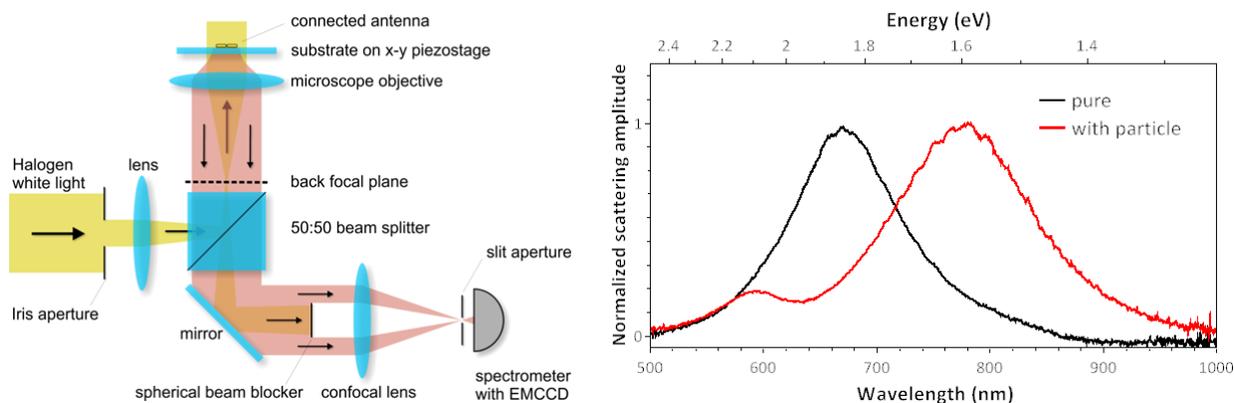

**Figure S2.1 | White-light dark field scattering measurements. Left:** Schematics of the setup and **Right:** Typical scattering spectra of an antenna without (pure) and with particle.

## Electrical Characterization

For electrical measurements the structures of interest were connected via their contact pads using a micromanipulator (DPP220, Cascade Microtech, Beaverton, USA) connected to a sourcemeter (Keithley 2601A, Keithley Instruments Inc., Cleveland, USA) while the other electrode was grounded. The sourcemeter was used to record I-V as well and I-t curves. Typical parameters were 0.1 NPLC (number of power line cycles) for the analog-to-digital converter, integration times of 100 ms and a current limit of 100 nA. For the I-V curves several measurements per voltage step were acquired (typically 5-10) to account for charging currents.

Connected antennas without a gap showed an overall resistance of ~200 Ω, from which we conclude that voltage drops in the supply line can be neglected. Antennas with a gap but without particle showed a resistance of more than 1 TΩ. This means that parasitic currents over the antenna gap can also be neglected.

# S3.  Electroluminescence Measurements

## Basic Setup and Software

Correct interpretation of electroluminescence spectra requires the knowledge of the overall spectral transfer function of the detection path (see Figure S3.1). The wavelength-dependent transfer functions of all individual elements (microscope objective, mirrors, confocal lens, spectrometer mirrors, grating) as well as the QE of the EMCCD were obtained from the respective manufacturers. The collection efficiency of the objective was assumed to be 50%. Using these data, corrected spectra and emitted photon numbers were determined.

Electrical excitation was realized using micromanipulators (DPP220, Cascade Microtech, Beaverton, USA) and a sourcemeter (Keithley 2601A, Keithley Instruments Inc., Cleveland, USA). In order to synchronize the electrical and optical measurements, the devices (sourcemeter, spectrometer, EMCCD) were controlled by a homebuilt LabVIEW program having a temporal accuracy of ~1 ms. Both, I-V and I-t modes were implemented and typically every 100 ms the current was recorded and every 100-500 ms spectra were acquired. For determining the total number of photons spectra were integrated numerically.

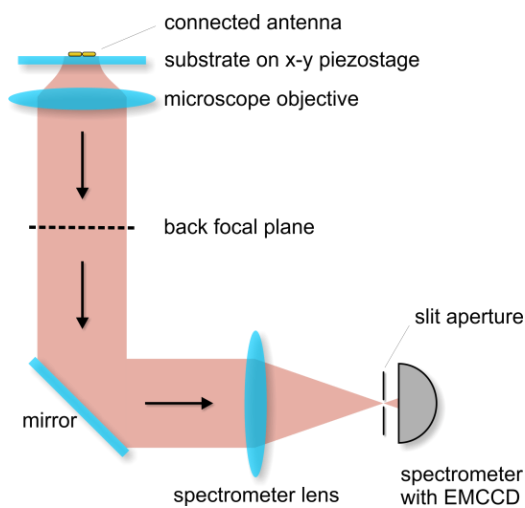

**Figure S3.1 | Schematics of the EL setup.** The beam block and splitter of the WL setup was removed (see Figure S2.1), the antenna was connected to a Keithley sourcemeter and a voltage was applied in order to record the EL characteristics.

## Gaussian Spot and Emission Pattern

For determining the FWHM of the electroluminescence emission spot, the spectrometer lens was replaced by a 1000-mm lens and the spectrometer was switched to mirror mode (c.f. Figure S3.2). In order to calibrate the resulting magnification a structure of known dimensions was imaged with the same arrangement.

To acquire electroluminescence emission patterns a 100-mm Betrand lens was placed in front of the EMCCD and the imaging lens was relocated to image the back focal plane (see Figure S3.2, right panel).

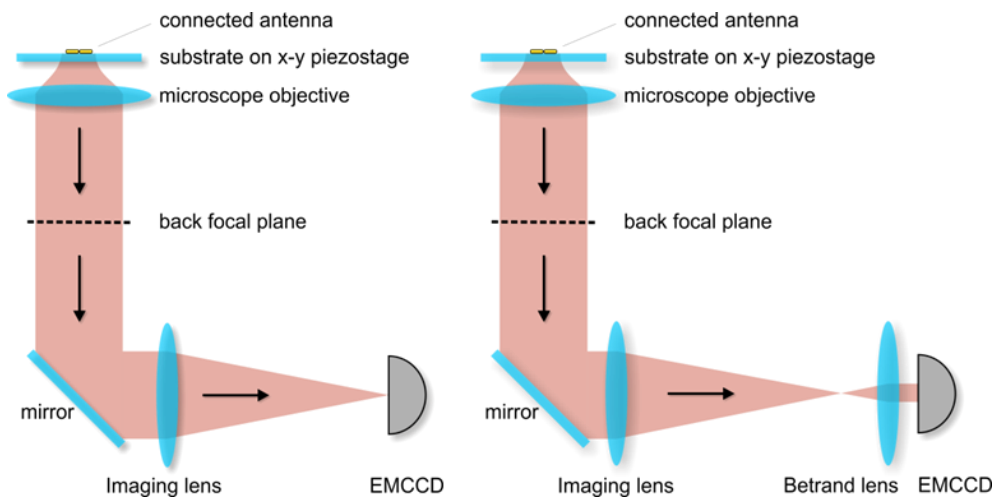

**Figure S3.2 | Schematics of further EL characterizations. Left:** Imaging of the emission spot. The spectrometer was switched to mirror mode, an imaging lens (f = 1000 mm) was installed and the spectrometer slit opened. **Right:** Acquisition of emission pattern. The imaging lens was displaced and a Betrand lens was installed in order to image the back focal plane of the microscope objective.

# S4. FDTD Calculations

## Determination of the Modes

Optical antennas are sensitive to their environment and it is expected that the optical properties for the loaded antenna are altered with respect to the isolated antenna, due to the particle situated in a high-field region within the gap. We investigate these effects by performing numerical finite-difference time-domain (FDTD) simulations (FDTD Solutions v8.7.3, Lumerical Solutions Inc., Vancouver, CAN).

The antenna is modeled as two rectangular particles with a width of 80 nm, a height of 50 nm and an arm length of 120 nm. The choice of approximating the antenna arms by rectangles is justified since scattering spectra are not sensitive to the exact shape[S4] and rectangles avoid the formation of hot spots due to stair casing. The antenna is situated on top of a $SiO_2$ (n=1.455) half-space. The dielectric constant of gold is modeled by an analytical fit to experimental data[S5]. A spherical particle with a diameter of 28 nm is placed in the 30 nm large gap between the antenna arms. A refinement mesh of 0.25 nm is placed over the gap region. The structure is excited by a total-field-scattered-field source and the scattering intensity is obtained by calculating the power-flow into the substrate direction.

The resulting scattering spectra are shown in Figure S4.1a. We compare an isolated antenna (bottom row) to three loaded antennas where the position of the particle with respect to the center is varied. The spectra are normalized, but all scattering intensities are of the same order of magnitude.

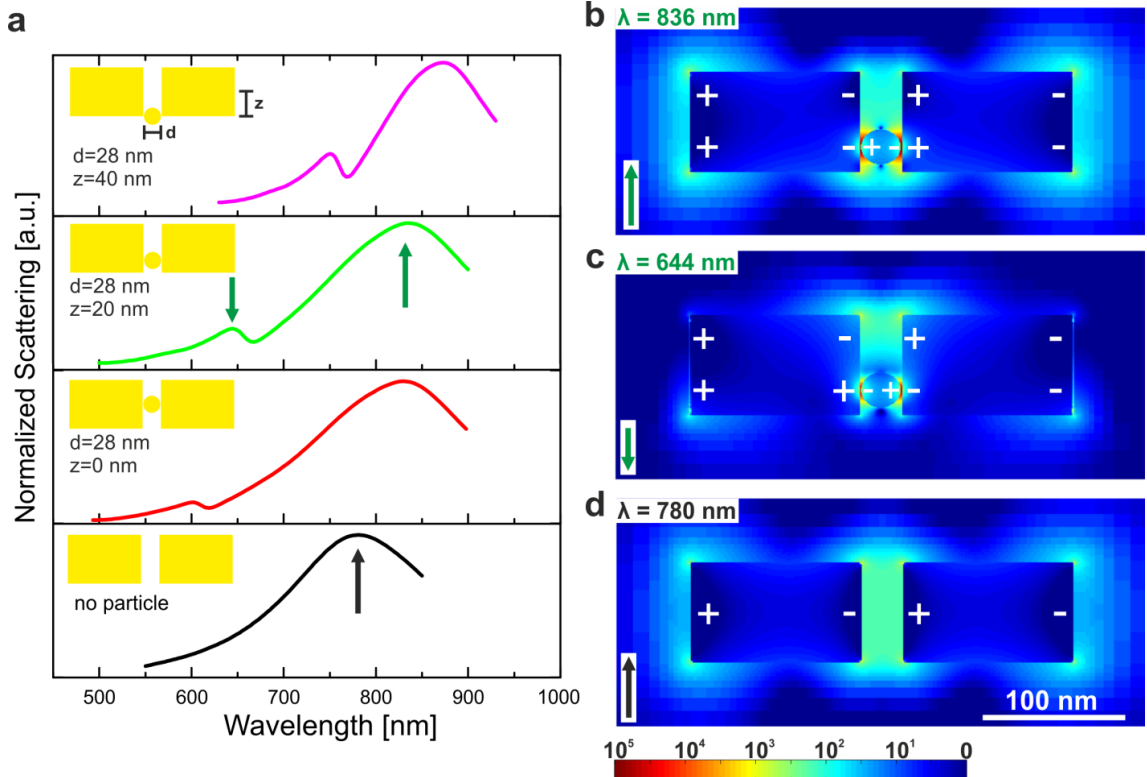

**Figure S4.1 | Influence of nanoparticle on optical properties. a** Scattering spectra obtained from FDTD simulations for different gold particle positions. **b-d** Near-Field intensity distributions in logarithmic color scale.

The antenna without particle shows a single resonance at 780 nm. The corresponding near-field intensity enhancement of the dipolar mode is about 500 and the intensity is concentrated between the antenna arms (Figure S4.1d).

On the other hand, the loaded antennas show two peaks, in agreement to loaded disk antennas[S6]. The spectral position of the peaks depends on the position of the particle and both peaks shift to larger wavelength when the particle is displaced from the center. The near-field intensity distributions of the two modes (Figure S4.1b,c) are calculated for a particle position 20 nm away from the center.

For the lower energy mode (Figure S4.1b) the charge distribution is very similar to the isolated antenna. The particle in the center is in phase with the antenna arms and high-field regions occur between the particle and both antenna arms. The intensity enhancement in this region is about $2\times10^5$, which is more than two orders of magnitude larger than for the antenna without particle. The resonance wavelength of the lower energy mode is shifted to larger wavelength with respect to the isolated antenna and its peak position strongly depends on the antenna length. Moreover, the peak position depends on the size of the particle and larger resonance wavelengths are observed for larger particles. Consequently, the particle seems to act as a polarizable material, which increases the surface charge accumulation at the gap.

For the higher energy mode (Figure S4.1c) the particle oscillates out of phase with respect to the charges at the outer side of the antenna arms. The charges at the inner side of the antenna, i.e. the gap region, are of opposite sign at the position of the particle compared to the position without particle. Consequently, a field minimum occurs at the very center of the antenna. The near-field intensity enhancement at the position between the polarized particle and both antenna arms is about $5\times10^4$, four times weaker than for the lower energy mode. The higher-energy mode strongly depends on the position of the particle as well as the width of the antenna. However, its peak position is only slightly affected by the length of the antenna.

The higher energy mode shows a lower field enhancement as well as drastically smaller radiation efficiency than the lower energy mode. Consequently, we can conclude that the light generation efficiency of the higher energy mode is significantly lower.

### Emission Pattern

We have numerically calculated the emission pattern of a single dipole situated above a glass halfspace by means of the FDTD method (Figure S4.2). The patterns were obtained by applying a far-field projection onto a hemisphere below the dipole in the direction of the substrate. As expected a clear two-lobed pattern is observed and emission is strongest at the critical angle. The emission pattern is in good qualitative agreement to the experimentally obtained characteristic of the radiated electroluminescence from the electrically-driven nanoantenna presented in the main text.

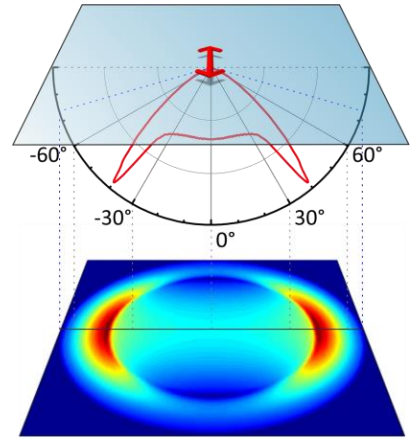

**Figure S4.2 | Fourier-space characteristic of dipole on a glass substrate.** Simulated radiation pattern of a single dipole.

# S5.  Fitting the FN-Characteristics

The current-voltage characteristics are obtained at ambient conditions using the setup described in section S2.

To obtain the width of the tunnel junction as well as the barrier height we modeled the current-voltage characteristic. In principle this is a difficult task but in order to gain some insight we applied the limited[S7] but analytical Simmons approximation[S8], which assumes a rectangular-shaped barrier of width $\Delta s$ and a mean barrier height $\varphi$. The dependence of current on voltage then reads as

$$I = C_S J_0 \{ \varphi \exp(-A\sqrt{\varphi}) - (\varphi + eV) \exp(-A\sqrt{\varphi + eV}) \},$$

where $J_0 = e/2\pi h(\beta \Delta s)^2$ and $A = 4\pi \beta \Delta s/h \sqrt{2m}$; h being Planck's constant and m the electron mass. The correction factor $\beta$ describes the deviation to the actual potential and is typically close to unity while the tunneling cross-section $C_S := w^2$ is in the order of 1 nm². We further take a leakage resistance (TΩ regime) as well as a current offset (pA regime) into account.

Images charges are known to round off the corners of the barrier and reduce its width. The potential barrier then depends also on the distance to the electrode $x$ and can be described as[S8,S9]

$$\varphi(x) = \varphi_0 - \frac{eVx}{s} - \frac{1.1\,\lambda s^2}{2x(s-x)} \;,$$

where $\lambda = e^2 \frac{\ln 2}{8\,\pi\varepsilon s}$. However, this classical description diverges for $x=0$ and $x=s$ and it is currently under debate if a classical description of image charges is suitable for nanometer-sized junctions[S10]. To estimate to what extend classical image charges play a role for the investigated junctions, we model the current-voltage characteristic with and without image charges. For doing that, $C_S$ was kept constant at 1 nm² and the leakage resistance as well as the current offset were set to 1 TΩ and 2 pA, respectively. A typical current-voltage characteristic and the corresponding fits are shown in a linear scale (Figure 0.1a) as well as in the Fowler-Nordheim representation (Figure 0.1b). We find best agreement to experimental data when image charges are taken into account. The open parameters are the width of tunnel junction as well as the barrier height. The obtained width is 1.6/1.3 nm and barrier height is 2.3/1.4 V for the model with/without image charges, respectively.

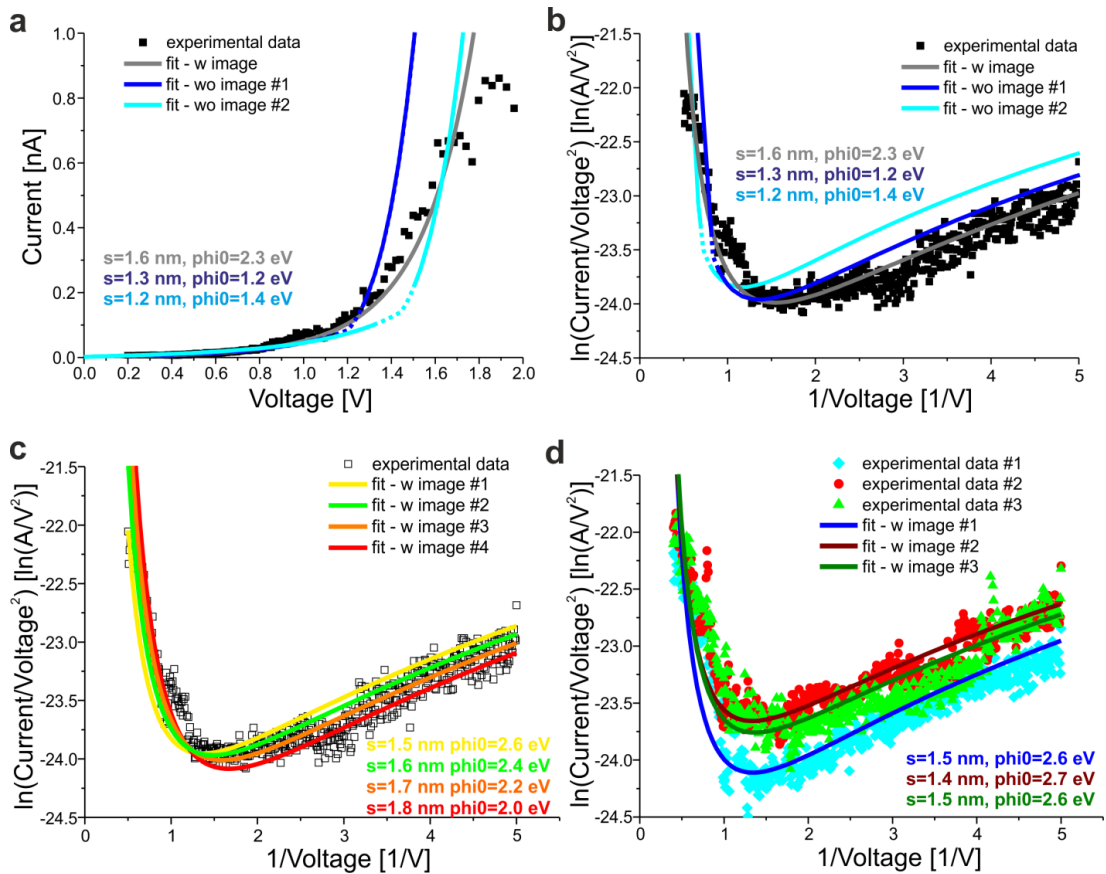

**Figure 0.1 | Electrical properties of nanoantennas with tunneling gaps**. **a**, Experimental current-voltage characteristic on a linear scale. Solid lines are non-linear fits to the data with (gray) and without (blue) the inclusion of image charge effects. **b**, Fowler-Nordheim representation of the data in (**a**). **c**, Fitting the same current-voltage characteristic with different fixed gap sizes. **d**, Three subsequently recorded current-voltage characteristics of the same structure with corresponding fits.

To estimate the uncertainties of gap size and barrier height the fitting procedure has been repeated for 4 different fixed gap sizes (1.5-1.8 nm) leaving the barrier height as the only open parameter. The results are shown in Figure 0.1c and reasonable fits are obtained for a gap size of 1.6 nm and a barrier height of 2.4 eV, as well as for a gap size of 1.7 nm and a barrier height of 2.2 eV. For a gap size of 1.5 nm and 1.8 nm deviations between the experimental data and the fit are clearly visible. Consequently, for this structure the resulting uncertainties for gap size are about 0.2 nm and 0.2 eV for the barrier height.

In order to test the stability of the investigated system three additional current-voltage characteristics have been measured for the same structure and the obtained data as well as corresponding fits are presented in Figure 0.1d. For the first measurement (blue rhombus) a gap size of 1.5 nm and a barrier height of 2.6 eV is obtained. The values from the second measurements (red circles) are 1.4 nm and 2.7 eV for gap size and barrier height respectively. The third measurement (green triangles) yields a gap size of 1.5 nm and a barrier height of 2.6 eV. These differences are within the above determined uncertainty and it can be concluded that the structure exhibits a stable tunneling junction.

# S6.  Time Stability

The tunneling current is very sensitive to the width of the junction. Movements of the particle within the gap or changes of the atomic configuration of the gap, e.g. by surface diffusion[S11] are expected to affect the tunneling current as well as the electroluminescence signal. Since all measurements are conducted at room temperature and ambient conditions, it is important to investigate the stability of the tunneling junction.

To this end, while a voltage of 2.0 V is applied, which corresponds to a field strength ~$1.5 \times 10^9$ V/m in the junction, the tunneling current is recorded simultaneously with electroluminescence spectra. By integration over the spectrum the total amount of emitted photons is obtained. In Figure S6.1a the tunneling current is plotted together with the number of radiated photons as a function of time. During the recorded 30 seconds the tunneling current increases slightly and shows some fluctuations. The emitted photons follow this trend. The electroluminescence spectrum during the same time interval is depicted in Figure S6.1b. The shape of the spectrum is very stable and no significant shifts in wavelengths are observed. Considering the fact that the experiments are performed at ambient conditions, the tunneling junctions show an astonishing good stability. This can be attributed to the molecular spacer layer around the spherical particle which is electro-statically bound to the antenna and stabilizes the gap. Even more stable junctions could be realized by exploiting molecules which covalently bind to the gold surface or by embedding the tunneling junction into an insulating material.

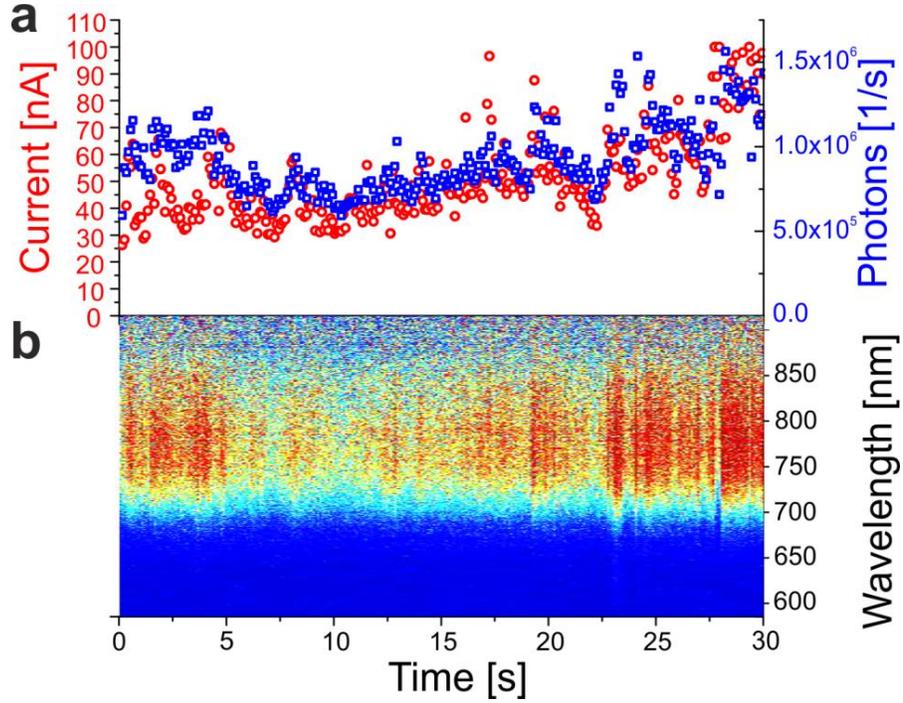

**Figure S6.1 | Time and spectral stability. a,** Tunneling current and number of emitted photons recorded simultaneously over 30 seconds. **b,** Spectrum of the electroluminescence during the same 30 seconds.

# S7. Current Power Model

As mentioned in the main text the spectrum of emitted photons *I(ω)* can be understood as a convolution of the current power spectrum and the plasmonic resonance, namely:

$$I(\omega) \propto A(\omega) \, C(\omega) = A(\omega)(eV - \hbar\omega)$$

where A(ω) is the emissivity of the antenna (i.e. the near- to far-field coupling due to the plasmonic antenna), ω the considered frequency and V the applied voltage. In Figure S7.1 the current power spectra are plotted for the voltages applied in Figure 2c (main text) and, additionally, for a higher voltage of 3.0 V. Furthermore, they are multiplied with a Lorentzian fit of the plasmonic resonance to model the light emission process. The resulting spectra agree very well with the experimental data (main text) and for 3.0 V the antenna scattering spectrum is also closely resembled.

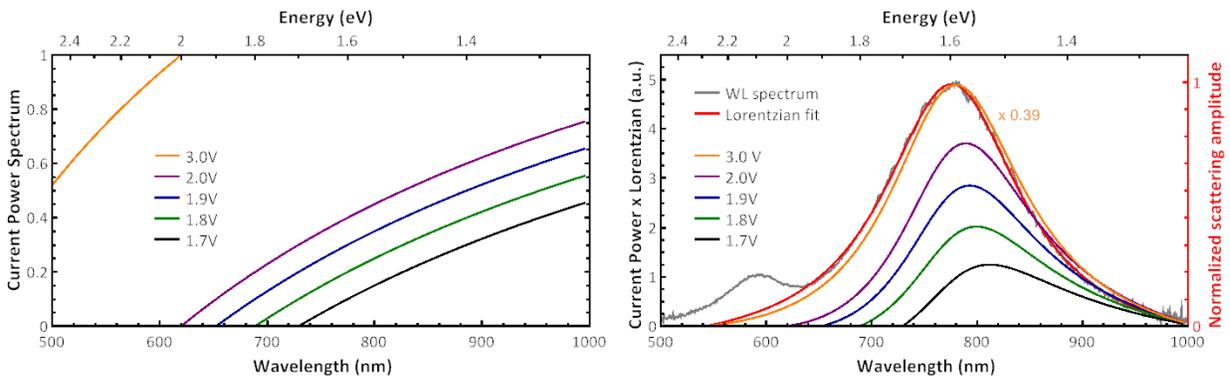

**Figure S7.1 | Current power model. Left:** Current power spectrum for five different voltages. **Right:** WL scattering spectrum from Figure 2c, Lorentzian fit and resulting emission spectra.

# S8.  Polarization Dependence

The polarization dependence was checked in two different configurations: (i) by simply placing a polarizer in the detection path and rotate it by 90° or (ii) by inserting a Wollaston prism that allows for simultaneous imaging of both polarizations onto the CCD. The later approach circumvents any uncertainties due to fluctuation of the electroluminescence and is therefore depicted in Figure S8.1. As a result we obtain a degree of polarization of about 11 limited by depolarization effects that occur in the high-NA objective.

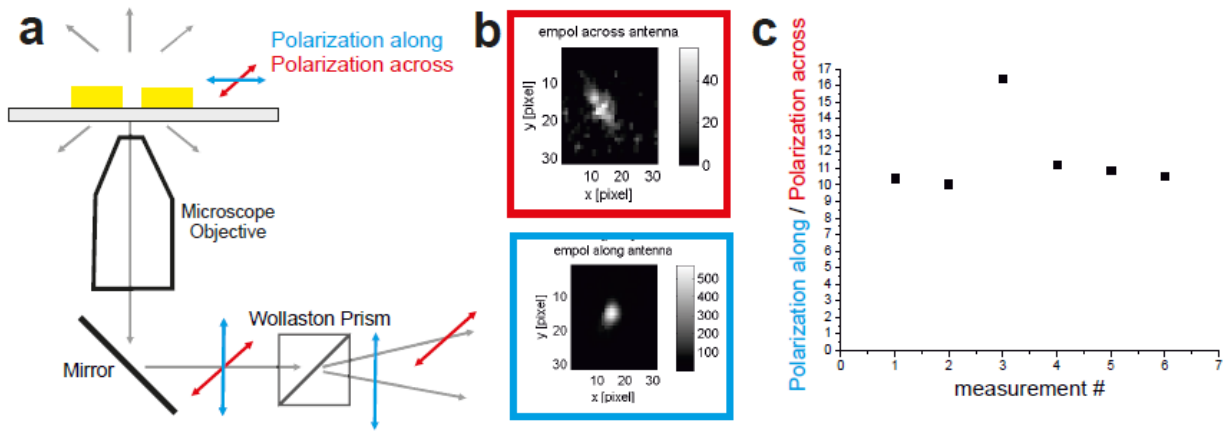

**Figure S8.1 | Polarization dependence of the electroluminescence. a,** Schematic sketch of the experimental setup. **b,** CCD images of the emitted light with polarization along the antenna (bottom, blue frame) as well as with polarizations across the antenna (top, red frame). Note the different scaling. **c,** Degree of polarization, i.e. polarization along divided by polarization across the antenna.